\documentclass[seceq]{ptptex}
\usepackage{wrapft}
\usepackage{graphicx}
\usepackage{bm}

\newcommand{\beq}{\begin{equation}}
\newcommand{\eeq}{\end{equation}}
\newcommand{\bea}{\begin{eqnarray}}
\newcommand{\eea}{\end{eqnarray}}
\renewcommand{\b}{\beta}
\renewcommand{\a}{\alpha}

\def\la{\mathrel{\mathpalette\fun <}}
\def\ga{\mathrel{\mathpalette\fun >}}
\def\fun#1#2{\lower3.6pt\vbox{\baselineskip0pt\lineskip.9pt
  \ialign{$\mathsurround=0pt#1\hfil##\hfil$\crcr#2\crcr\sim\crcr}}}
\newcommand{\bfi}[1]{\mbox{\boldmath $#1$}}
\newcommand{\bfis}[1]{\mbox{\boldmath ${\scriptstyle #1}$}}
\newcommand{\vb}{{\bfi b}}
\newcommand{\vk}{{\bfi k}}

\newcommand{\vq}{{\bfi q}}

\newcommand{\vL}{{\bfi L}}
\newcommand{\vrr}{{\bfi r}}
\newcommand{\vrR}{{\bfi R}}
\newcommand{\vx}{{\bfi x}}
\newcommand{\vy}{{\bfi y}}
\newcommand{\vib}{{\bfis b}}

\newcommand{\vir}{{\bfis r}}
\newcommand{\viR}{{\bfis R}}

\newcommand{\viq}{{\bfis q}}

\title{
A New Glauber Theory based on Multiple Scattering Theory
}

\author{
Masanobu {\sc Yahiro},
\footnote{E-mail:~yahiro@phys.kyushu-u.ac.jp}
Kosho {\sc Minomo},
\footnote{E-mail:~minomo@phys.kyushu-u.ac.jp}
Kazuyuki {\sc Ogata},
\footnote{E-mail:~ogata@phys.kyushu-u.ac.jp}
\\
and Mitsuji {\sc Kawai}
\footnote{E-mail:~kawa2scp@kyudai.jp}
}

\inst{
Department of Physics, Kyushu University, Fukuoka 812-8581,
Japan
}

\recdate{
\today
}

\abst{
Glauber theory for nucleus-nucleus scattering at 
high incident energies is reformulated so as to also be applicable 
to the scattering at intermediate energies. 
We test the validity of the eikonal and adiabatic approximations used 
in the formulation, and 
discuss the relationship 
between the present theory and the conventional Glauber 
calculations using either the empirical nucleon-nucleon profile function or 
the modified one including the in-medium effect. 
}
\begin{document}
\maketitle

\section{Introduction}
\label{Introduction}

Experiments with radioactive beams of unstable nuclei have opened
new frontiers in nuclear physics.
New features of unstable nuclei
such as a halo structure were revealed;
see, for example, Ref. \citen{Tanihata}.
The Glauber theory \cite{Glauber} has widely been used
as a powerful tool for studies of reactions with unstable nuclei
observed at intermediate energies
such as 50 -- 800 MeV/nucleon \cite{Tostevin,Ogawa01}.
It was reported \cite{Xiang,Horiuchi:2006ga},
however, that
some modifications of the nucleon-nucleon (NN)
scattering profile functions in the eikonal approximation were
necessary in order to reproduce
the data at energies of less than 500 MeV/nucleon \cite{Xiang,Horiuchi:2006ga}.
Such phenomenological modifications obviously require theoretical
foundations. In this paper, we address this problem.

The Glauber theory describes the scattering of two nuclei P and A
as collisions of all nucleons in P with those in A.
The theory starts with the many-body Schr\"odinger equation
\beq
\big[ E-K-h_{\rm P}-h_{\rm A}-V \big] \Psi=0 , \quad
\label{original-Schrodinger}
\eeq
where $V=\sum_{i \in {\rm P},j \in {\rm A}} v_{ij}$ with $v_{ij}$
the NN interaction potential,
$E$ is the energy of the total system,
$K$ is the kinetic energy operator of relative motion
between P and A, and
$h_{\rm P}$ ($h_{\rm A}$) is the internal Hamiltonian of P (A).
Assuming the adiabatic approximation for the internal motion of P and A and
the eikonal approximation,
the theory gives
the scattering amplitude
at high energies and small scattering angles as
\bea
f_{\b\a}= \frac{ik}{2\pi }
\int d\vb\;
e^{i \viq \cdot \vib } \;
\langle \Phi_{\b} \vert
1-\prod_{i,j}(1-\Gamma_{\rm {NN}}(\vb_{ij}))
\vert \Phi_{\a} \rangle ,
\label{fPA}
\eea
where
$\hbar \vk$ ($\hbar \vq$) is the initial (transferred) momentum,
$\vb$ is the component of the relative
coordinate $\vrR$  between the centers of mass
of P and A perpendicular to $\vk$,
$\Phi_{\a}$ ($\Phi_{\b}$) is the antisymmetrized internal wave function
of the initial (final) channel,
and $\vrr_{ij}=(\vb_{ij},z_{ij})$
is the displacement of $i$ from $j$ with
$\vb_{ij}$ ($z_{ij}$) the component of $\vrr_{ij}$
perpendicular (parallel) to $\vk$.
Here, $\Gamma_{\rm {NN}}(\vb_{ij})$ is the profile
function of the scattering of nucleon $i$ in P and nucleon 
$j$ in A given by
\bea
\Gamma_{\rm {NN}}(\vb_{ij})=1-{\rm exp}
\Big[
-{i \over \hbar v_{\rm rel}} \int_{-\infty}^{\infty}
dz_{ij}\; v_{ij}(\vrr_{ij})
\Big] ,
\label{GammaNN}
\eea
where $v_{\rm rel}$ is the relative velocity.

If the eikonal approximation is valid for NN scattering in free space,
$\Gamma_{\rm {NN}}(\vb_{ij})$ in (\ref{GammaNN}) 
should agree with the Fourier
transform $\Gamma^{\rm em}_{\rm {NN}}$ 
of the NN scattering amplitude $f_{\rm NN}$ 
determined from the data on NN scattering in free space,
\beq
\Gamma^{\rm em}_{\rm {NN}}(\vb_{ij})=\frac{-i}{2\pi k_{ij}}
\int e^{-i\vq_{ij} \cdot \vb_{ij}}
f_{\rm NN}(\vq_{ij})
d\vq_{ij},
\label{EqMK1-5}
\eeq
where $\vk_{ij}$ ($\vq_{ij}$) is the initial (transferred)
momentum of relative motion between the two nucleons $i$ and $j$
in free space and
the 2-dimensional integration is over the components of
$\vq_{ij}$ perpendicular to $\vk_{ij}$.
Instead of using $\Gamma_{\rm {NN}}(\vb_{ij})$, 
$\Gamma^{\rm em}_{\rm {NN}}(\vb_{ij})$ 
has been used customarily, since first introduced in Ref.~\citen{GM}.
However, it should be noted that
the replacement of $\Gamma_{\rm {NN}}(\vb_{ij})$ by
$\Gamma^{\rm em}_{\rm {NN}}(\vb_{ij})$ in (\ref{fPA}) 
is correct only if the eikonal approximation is valid for the NN 
scattering in free space, 
since (\ref{GammaNN}) has already been used in the derivation of (\ref{fPA}).

Substituting (\ref{EqMK1-5}) into (\ref{fPA}), one can calculate the
nucleus-nucleus (PA) scattering amplitude $f_{\beta\alpha}$. 
In Ref. \citen{GM}, $f_{\rm NN}(\vq_{ij})$ was given,
neglecting the Coulomb effects, in the form
\beq
f_{\rm NN}(\vq_{ij})=
f_{\rm NN}(0)
e^{-\beta^2 \vq^2_{ij}/2}
\label{EqMK1-6}
\eeq
with
\beq
f_{\rm NN}(0)=
(i+\alpha)
k_{ij} \sigma/4,
\label{EqMK1-7}
\eeq
where $\sigma$ is the NN total cross section,
and the constants $\alpha$ and $\beta$ were derived from the experimental
data \cite{GM}. This prescription works well at high energies.
As already mentioned, however, (\ref{EqMK1-7}) had to be modified
at energies lower than 500 MeV. The parameters $\alpha$ and $\beta$
had to be adjusted away from those made to fit the PA scattering data. 
This cast doubt
on the prescription, in particular on
the validity of the eikonal approximation to NN scattering.
In fact, the condition for the validity of
the eikonal approximation to NN scattering is
\bea
\vert v_{ij}/E_{ij} \vert \ll 1, \quad  k_{ij}a \gg 1 \;,
\label{condition}
\eea
where $E_{ij}$  and $k_{ij}$ are, respectively,
the kinetic energy and the wave number
of the relative motion of $i$ and $j$, and
$a$ is the width of the region(s) in which $v_{ij}$ changes rapidly.
However, since $v_{ij}$ has a strong
short-range repulsive core, for example
$v_{ij} \sim 2000$ MeV at $r_{ij}=0$ in the case of the realistic
NN potential AV18 \cite{Wiringa},
it is obvious that the first condition of (\ref{condition}) is not 
satisfied.

In order, therefore, to examine the accuracy
of the eikonal approximation to NN scattering,
we calculate the NN scattering amplitude
at the laboratory energy $E_{\rm {NN}}=300$ MeV
with the eikonal approximation and compare the result with the exact one.
Figure 1 shows the on-shell NN scattering amplitude $f_{\rm NN}(\vq_{ij})$.
For simplicity,
we take only the central part of the realistic NN potential AV18
for the triplet-even state.
The solid (dashed) and dotted (dash-dotted) lines show, 
respectively,
the real and imaginary parts of the resulting scattering amplitude
of the exact (eikonal) calculation.
The eikonal amplitude deviates considerably from the exact one,
even at small $\vq$, for both the real and imaginary parts.
In particular, the deviation is serious on the imaginary part.
This means that the Glauber theory can not accurately predict the reaction
cross sections of nucleon-nucleus (NA) and nucleus-nucleus (PA)
scattering.
This test clearly shows that
the use of (\ref{EqMK1-5}) in (\ref{fPA})
is inaccurate.

\begin{figure}[htb]
\centerline{\includegraphics[width=0.5\textwidth]{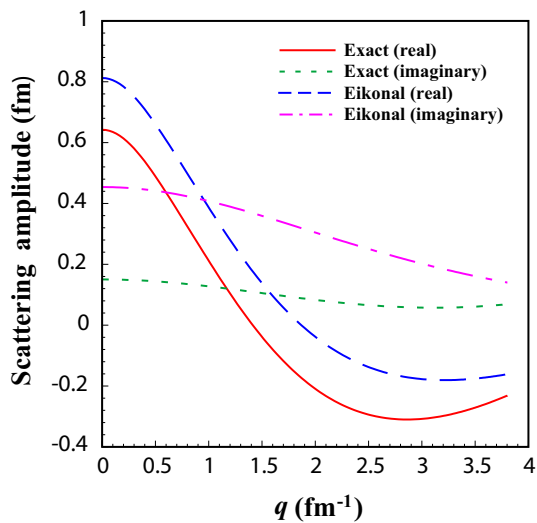}}
\caption{\label{fig:fcomp}
Test of eikonal approximation for 
on-shell NN scattering amplitude $f_{\rm {NN}}(\vq)$
at the laboratory energy $E_{\rm {NN}}=300$ MeV. 
As the NN interaction, 
the central part of AV18 for the triplet-even state 
is taken.
}
\end{figure}

A systematic way of making non-eikonal corrections was
proposed in Ref. \citen{Wallace} for high-energy potential scattering.
In the method, the transition matrix is expanded in a power series of
$\delta=|v_0|/(k \hbar v_{\rm rel})$,
where $v_0$ is a typical potential strength.
The method, however, has not been applied to
intermediate-energy NN scattering because $\delta >1$ in such a case.

For the NA  scattering mainly at high incident energies,
some methods for treating non-eikonal effects
have been proposed \cite{Wong, Wallace-review},
but they require much more difficult and/or
complicated calculations than the Glauber approximation.
For example, Wong and Young introduced the pseudopotential $v_{ij}'$
that reproduces the empirical
NN scattering amplitude
in the eikonal approximation
and estimated the corrections to
the Glauber amplitude when $v_{ij}'$ is used
instead of $v_{ij}$~\cite{Wong}.
However, it is not easy to evaluate the corrections
induced by the difference $v_{ij}-v_{ij}'$,
since $|v_{ij}'| \ll | v_{ij}|$ and, therefore,
$v_{ij}-v_{ij}'$ is much larger than $v_{ij}'$.

It was proposed in Ref. \citen{Xiang} that
$f_{\rm {NN}}(\vk', \vk)$ is modified from the empirical
NN scattering amplitude in free space to that calculated with
the Brueckner $g$-matrix.
Although this seems a reasonable proposition,
it is obvious that the theoretical foundations of such a phenomenological
procedure need be examined.

In this paper, we propose an accurate and practical method
of treating the intermediate-energy PA scattering
within the framework of the Glauber theory.
For this purpose, we use
the effective NN interaction $\tau$ that describes the NN collisions
in PA scattering instead of the bare potential $v_{ij}$.
We introduce this effective NN interaction in the Glauber theory
on the basis of the multiple scattering
theory (MST) of Watson \cite{Watson} using the formalism
of Kerman, McManus, and Theler (KMT) \cite{KMT}.
It turns out that the $\tau$ satisfies the condition (\ref{condition})
better than $v_{ij}$ and can even take account of the modification of
the NN interaction in the nuclear medium that has long been thought
to be necessary for reactions at low and intermediate 
incident energies \cite{Xiang}.

For the high-energy nucleon-nucleus scattering in which 
the Glauber approximation is good and, therefore, 
(\ref{GammaNN}) is accurate, 
the relationship between the Glauber theory 
and MST was investigated 
in detail~\cite{Glauber-foundation, Wallace-review}.
These studies, however, were mostly concerned with 
the cancellation in the Watson expansion between the reflection terms and 
the off-pole contribution of nonreflective terms~\cite{Glauber-foundation}. 
This is not addressed in the present paper. 
We present a method of going beyond
the ordinary Glauber theory by calculating all the
Watson series using the Glauber approximation.

This paper is organized as follows.
In \S~\ref{Formulation}, we present a new version
of the Glauber theory based on the effective NN interaction $\tau$.
In \S~\ref{Discussions}, some important points are
discussed.
In \S~\ref{Effective interaction},
we approximate $\tau$ as a two-body operator such
as the transition matrix of NN scattering in free space or
the Brueckner $g$-matrix, and we presents a way of localizing the
two-body operator.  
In \S~\ref{Validity}, we test 
the eikonal and adiabatic approximations used in the present formalism. 
Such a test is not feasible for PA scattering 
consisting of an infinite series of multiple NN scatterings. 
The test is then made for a single NN collision in PA scattering, 
which is an essential building block of PA scattering. 
In \S~\ref{Relation}, we discuss 
the relationship between the present theory and 
the conventional Glauber calculations
using either the empirical NN profile function (\ref{EqMK1-5}) or the 
modified one including the in-medium effects \cite{Xiang}.
Section \ref{Summary} is devoted to a summary.

\section{Formulation}
\label{Formulation}

\subsection{Multiple scattering theory for nucleus-nucleus scattering}
\label{MST}

The original KMT formalism \cite{KMT}
is for nucleon-nucleus (NA) scattering.
We first extend it to the case of
nucleus-nucleus (PA) scattering between P and A.
The transition matrix of PA scattering is given by
$T = V(1 + G_0T)$ with
\bea
G_0=\frac{\cal P}{E-K-h_{\rm P}-h_{\rm A}+i\epsilon} ,
\label{T-AA}
\eea
where
${\cal P}={\cal P_{\rm P}}{\cal P_{\rm A}}$ with
${\cal P_{\rm P}}$ (${\cal P_{\rm A}}$)
the projection operator onto the space
of antisymmetrized wave functions of P (A).

The transition matrix $T$ is
given by
\bea
T=\sum_{i \in {\rm P}, j \in {\rm A}} T_{ij} \;,
\label{Tij}
\eea
where $T_{ij}$ satisfy a set of coupled equations
\bea
T_{ij}=\tau_{ij}+\tau_{ij} G_0 \Big\{
\sum_{m \in {\rm P}, n \in {\rm A}} T_{mn}-T_{ij}
\Big\}
\;,
\label{multiple}
\eea
where
\bea
\tau_{ij}=v_{ij}(1+G_0\tau_{ij})
\label{tau}
\eea
is an operator that
describes the scattering of nucleons $i$ in P and $j$ in A.
The potential $v_{ij}$, in general, contains operators
acting on the spins and isospins of $i$ and $j$, which we suppress for
the simplicity of notation.

The proof of (\ref{multiple}) is as follows. We define $X$ as
$X=\sum_{i \in {\rm P}, j \in {\rm A}} T_{ij}$.
Equation (\ref{multiple})
is then reduced to $T_{ij}=(1+\tau_{ij} G_0)^{-1}\tau_{ij}(1+G_0 X)$.
Identifying $(1+\tau_{ij} G_0)^{-1}\tau_{ij}$ with $v_{ij}$ in the equation
and summing $T_{ij}$ over $i$ and $j$, one can obtain $X=V(1+G_0 X)$.
We then find that $X=T$.

Because of the antisymmetry of the nuclear wave functions,
which is maintained by ${\cal P}$, the matrix elements of
the operators $T_{ij}$ and $\tau_{ij}$ are independent of
the labels $i$ and $j$, so that (\ref{multiple}) can be written as
\bea
\theta=\tau+(Y-1)\tau G_0 \theta \;,
\label{theta}
\eea
where $\theta=T_{ij}$ and 
$Y=N_{\rm P} \times N_{\rm A}$, with
$N_{\rm P}$ ($N_{\rm A}$) the nucleon number of P (A).
Multiplying (\ref{theta}) by $Y-1$, one obtains
\bea
T'=U(1+G_0T') \;,
\label{T'}
\eea
where
\bea
T'=(Y-1)\theta, \quad U=(Y-1)\tau \;.
\eea
Then, $T$ is obtained from $T'$ as
\bea
T=\frac{Y}{Y-1}T' \;.
\label{T-def}
\eea

Equations (\ref{T'}) and (\ref{T-def}) constitute
the final results of the extended KMT.
The antisymmetrization between the incident nucleons in P and 
the target nucleons in A has been neglected so far.
It was shown \cite{Takeda,Picklesimer}, however,
that it can be taken care of using $\tau$ which is symmetrical with respect to
the exchange of the colliding nucleons.

\subsection{Glauber approximation}
\label{Glauber-approximation}

We proceed to the calculation of the matrix elements $T'_{\beta \alpha}$
for the transition $\alpha \to \beta$
using the Glauber approximation.
We first define the wave matrix $\Omega_{\a}^{(+)}$
that gives the wave function with an incident wave in channel $\alpha$
by
${\hat\Psi}_{\a}^{(+)}
=\Omega_{\a}^{(+)}\Phi_{\a}\phi_{\a}$,
where $\phi_{\a} = (2\pi)^{-3/2} \exp[i \vk \cdot \vrR]$.
The wave function ${\hat \Psi}_{\a}^{(+)}$ satisfies
\bea
(K+h_{\rm P}+h_{\rm A}+U-E){\hat \Psi}_{\a}^{(+)}=0  \;.
\label{schrodinger}
\eea
The matrix elements of $T'$ for the transition $\a \rightarrow \b$ 
are then given
by $T'_{\b\a}=\langle \Phi_{\b}\phi_{\b}|U|{\hat \Psi}_{\a}^{(+)}\rangle$.
In the present formalism, $\Phi_{\a}$ is an eigenstate of 
the internal Hamiltonian $h_{\rm P}+h_{\rm A}$ with 
the realistic NN interaction $v_{ij}$ . 
In actual calculations, however, the $v_{ij}$ is often replaced 
by an effective interaction that usually 
has a weak repulsion. The short-range correlation neglected by the 
replacement may not be important at forward PA scattering with 
small $q$.

The potential $U$ can be written in the form
\beq
U=\frac{Y-1}{Y}\sum_{i \in {\rm P},\; j \in {\rm A} } \tau_{ij}
\label{Utauij}
\eeq
because of the total antisymmetry of the wave functions of P and A.
By definition (\ref{tau}), $\tau_{ij}$ is a many-body operator
acting on all the nucleons of the total system.
We assume, however, as commonly performed in practical applications of MST
and discussed in detail in \S~\ref{Effective interaction},
that $\tau_{ij}$ can be well approximated by a two-body operator
acting only on $i$ and $j$ and in coordinate representation by
a local potential depending only on $\vrr_{ij}$.

Equation (\ref{schrodinger}) then has the same form
as the original Schr\"odinger equation (\ref{original-Schrodinger})
for the wave function of
the total system except that $V$ is replaced by $U$.
The Glauber approximation can,
therefore, be applied to (\ref{schrodinger}) if the
same conditions are satisfied,
i.e., the adiabatic approximation to the internal motion of P and A is used 
to approximate $h_{\rm P}$ and $h_{\rm A}$ by their ground-state energies,
and the conditions for the eikonal approximation, 
\bea
\vert (Y-1)\tau/E \vert \ll 1 \;, \quad  k a_{\tau} \gg 1 \;,
\label{condition-U}
\eea
are satisfied, 
where $a_{\tau}$ is the range of the region in which $\tau$ changes rapidly.
$\tau_{ij}$ has a much weaker dependence on $\vrr_{ij}$ than $v_{ij}$,
as shown in \S~\ref{Effective interaction}. Hence,
it satisfies the conditions in (\ref{condition-U}) better than 
the $v_{ij}$ of the original Glauber theory.

We denote the center of mass of P (A) by $\vrR_{\rm P}$ ($\vrR_{\rm A}$)
and the coordinate of nucleon $i$ ($j$) in P (A) by $\vrr_i$ ($\vrr_i$).
Then, $\vrR  = \vrR_{\rm P} - \vrR_{\rm A}$
is the relative coordinate of P from A
by which the scattering is described, and $\vx_i = \vrr_i - \vrR_{\rm P}$
($\vy_j = \vrr_j - \vrR_{\rm A}$) is the intrinsic coordinate of
$i$ ($j$) in P (A).
With this notation,
$U = U(\vrR,\xi)$ where $\xi = (\{\vx_i \}_{\rm P} ,\{\vy_j \}_{\rm A} )$.
For simplicity of notation, however, the intrinsic coordinate $\xi$
is suppressed in the following.

In the Glauber approximation, $T'_{\b\a}$ are given by
\bea
T'_{\b\a}=C
\int d\vb \; e^{i \viq \cdot \vib}
\langle \Phi_{\b}| \Gamma_{U}(\vb) |\Phi_{\a}\rangle \;,
\eea
where $C=-i \hbar^2 k/((2\pi)^3 \mu_{\a})$ with
$\mu_{\a}$ the reduced mass in the initial channel $\a$,
and the profile function of PA scattering is given by
\bea
&&\Gamma_U(\vb)=1-{\rm exp}[i\chi_U(\vb)] , \\
&&\chi_U(\vb)=-{1 \over \hbar v_{\rm rel}}
\int_{-\infty}^{\infty} dz\; U(z,\vb),
\label{chiU}
\eea
where $v_{\rm rel}=\hbar k/\mu_{\a}$.
Using (\ref{Utauij}), one can rewrite the phase shift
function $\chi_U(\vb)$ as
\bea
\chi_U(\vb)&=&\frac{Y-1}{Y}\sum_{i \in {\rm P},\; j \in {\rm A} }
\chi_{\rm {NN}}^{({\rm eff})}(\vb_{ij}) \;,
\eea
where
\bea
\chi_{\rm {NN}}^{({\rm eff})}(\vb_{ij})
&=&-{1 \over \hbar v_{\rm rel}} \int_{-\infty}^{\infty}
dz_{ij} \; \tau(z_{ij},\vb_{ij}) \;
\label{chi-NN}
\eea
is the phase shift function of NN scattering by effective interaction $\tau$.
The transition matrix elements are then given by
\bea
&&T_{\b\a} = \frac{Y}{Y-1} C
\int d\vb\;
e^{i \viq \cdot \vib}
\langle \Phi_{\b} \vert \Gamma_U(\vb) \vert \Phi_{\a} \rangle ,
\label{T}
\\
&&\Gamma_U(\vb) = 1-
\prod_{i=1}^{P}\prod_{j=1}^{A}(1-\Gamma_{\rm {NN}}^{({\rm eff})}(\vb_{ij})) ,
\label{Gamma-U}
\\
&&\Gamma_{\rm {NN}}^{({\rm eff})}(\vb_{ij}) =
1-{\rm exp}[\frac{Y-1}{Y}i \chi_{\rm {NN}}^{({\rm eff})}(\vb_{ij})] .
\label{GammaNN-eff}
\eea
The scattering amplitude $f_{\b\a}$
and the cross section $\sigma_{\b\a}$ are given by
\bea
f_{\b\a}=-\frac{(2\pi)^2\mu_{\b}}{\hbar^2} T_{\b\a} \quad {\rm and} \quad
\frac{d\sigma_{\b\a}}{d\Omega_{\b}}=|f_{\b\a}|^2 \;,
\label{fbetaalpha}
\eea
respectively.
Equations (\ref{T}) -- (\ref{GammaNN-eff}) with (\ref{chi-NN})
constitute the principal result of this paper, 
where the conventional form of the 
Glauber theory with (\ref{fPA}) and (\ref{EqMK1-5})
is reformulated using the effective interaction $\tau$.

Most PA collisions satisfy $Y \gg 1$.
The factors $(Y-1)/Y$ and $Y/(Y-1)$ in $T$ only 
have appreciable effects on 
collisions between very light nuclei. 
For the other collisions, 
the resulting scattering amplitude has the form of
the original Glauber theory except that
the bare NN interaction $v$ is replaced by $\tau$.

\subsection{Elastic scattering}
\label{Elastic-scattering}

For elastic scattering, where $\a=\b$, one can rewrite (\ref{T})
in the form
\bea
&&T_{\a\a} = \frac{Y}{Y-1} C
\int d\vb\;
e^{i \viq \cdot \vib}
(1-e^{i \chi_{\rm opt}(\vib)}) \;
\label{Topt}
\eea
with
\bea
&&\chi_{\rm opt}(\vb)= -i
\ln{\langle \Phi_{\a}| {\rm exp}[i\chi_U(\vb)] |\Phi_{\a}\rangle } \;.
\label{chiopt}
\eea
Since $\chi_{\rm opt}(\vb)$ is a function of $\vb$, there must be a
one-body potential $U_{\rm opt}(z,\vb)$ such that
\bea
\chi_{\rm opt}(\vb)=-{1 \over \hbar v_{\rm rel}} \int_{-\infty}^{\infty} dz\;
U_{\rm opt}(z,\vb) \;.
\label{Uopt}
\eea
Since $U_{\rm opt}(z,\vb)$ is a one-body potential that describes
the elastic scattering, it is the optical potential.
This definition of $U_{\rm opt}(z,\vb)$ differs from the ordinary one by
the overall factor $Y/(Y-1)$ in $T$,
which is, as has already been mentioned, 
nearly equal to 1 except for scattering
between very light nuclei.
The potential $U_{\rm opt}(z,\vb)$  is obtained as
the solution of (\ref{Uopt}):
\bea
U_{\rm opt}(r) = \frac{\hbar v_{\rm rel}}{\pi}\frac{1}{r}
\frac{d}{dr} \int_{r}^{\infty}
\frac{\chi_{\rm opt}(b)b db}{(b^2-r^2)^{1/2}} \;.
\label{inverse}
\eea
One can obtain $T_{\a\a}$ by
solving the Schr\"odinger equation using 
the optical potential $U_{\rm opt}(\vrr)$.

In the case of $|\chi_U(\vb)| < 1$,
${\rm exp}[i\chi_U(\vb)]$ in (\ref{chiopt})
can be expanded
in powers of $\chi_U(\vb)$. This
leads to the cumulant expansion of $\chi_{\rm opt}(\vb)$ \cite{Glauber}.
If one retains only the lowest-order term, one
obtains the optical limit,
$
\chi_{\rm opt}^{(0)}(\vb)
= \langle \Phi_{\a}| \chi_{U}(\vb) |\Phi_{\a}\rangle ,
$
and
\bea
U_{\rm opt}^{(0)}(\vrR)
= \frac{Y-1}{Y}
\langle \Phi_{\a}| \sum_{i \in {\rm P},\; j \in {\rm A} }
\tau(z_{ij},\vb_{ij})
|\Phi_{\a}\rangle \;,
\label{U-opt}
\eea
where the superscript (0) stands for the optical limit.
Equation (\ref{inverse}) includes all orders of the cumulant expansion,
although the calculation is not easy.

The optical theorem
$\sigma_{\rm {tot}}=4\pi{\rm Im}f_{\a\a}(\vk,\vk)/k$
for the total cross section $\sigma_{\rm {tot}}$ yields
from (\ref{fbetaalpha}) and (\ref{Topt})
\bea
\sigma_{\rm {tot}} =\frac{2Y}{Y-1}\int d\vb \;
{\rm Re}
\langle \Phi_{\a}|\Gamma_U(\vb)|\Phi_{\a}\rangle \;.
\eea
The angle-integrated cross section of elastic scattering is
\bea
\sigma_{\rm el}&&=\int d\Omega_{k'} \; |f_{\a\a}(\vk',\vk)|^2
\nonumber
\\
&&\approx \Big(\frac{Y}{Y-1}\Big)^2
\int d\vb \;
| \langle \Phi_{\a}|\Gamma_U(\vb)|\Phi_{\a}\rangle |^2 \;.
\label{Xsec-el}
\eea
When the elastic scattering is concentrated in the forward direction,
$d\Omega_{k'}$ is nearly on a plane perpendicular to the direction of $\vk$,
so that  $d\Omega_{k'} \approx d^{2}{\vk'}/{k'}^2$,
where $d^{2}{\vk'}$ is an area element on that plane \cite{Glauber}.
The reaction cross section $\sigma_{\rm reac}$ is then
obtained as
\bea
\sigma_{\rm reac}=\sigma_{\rm tot}-\sigma_{\rm el}=
\int d\vb \;
\Big( 1-|Z|^2 \Big) \;,
\label{sigma-reac-AA}
\eea
where $Z=- 1/(Y-1)+Y/(Y-1) \cdot S'_{\a\a}$ with
\bea
S'_{\a\a}=\langle \Phi_{\a}|
{\rm exp}\Big[
\frac{Y-1}{Y}\sum_{i \in P, j\in A} i \chi_{\rm {NN}}^{({\rm eff})}(\vb_{ij})
\Big]
|\Phi_{\a}\rangle .
\eea
In the case of $Y \gg 1$, which most collisions satisfy,
the reaction cross section of (\ref{sigma-reac-AA})
is reduced to that in the ordinary
Glauber method with $v$ replaced by $\tau$.

\section{Discussion}
\label{Discussions}

\subsection{Effective interaction}
\label{Effective interaction}

The effective interaction $\tau$ is
the key to not only the present formalism but also to 
various theories of direct reactions.
The effective interaction $\tau$ is a many-body operator,
because Eq. (\ref{tau}) for $\tau$ includes the many-body Green's function
$G_0$, which depends on all the internal coordinates of P and A.
There is a long history of development of
various approximations for $\tau$.
Those widely used in many practical applications are to simply replace $\tau$
by a two-body operator. In the impulse approximation, $\tau$ is approximated
by $t$, the transition matrix of NN scattering in free space
\cite{Love-Franey}.
This has been successful for describing NA scattering 
at high energies of $E_{\rm NA} \ga 500$~MeV \cite{KMT,Ray92,Arellano}.
At lower energies of $E_{\rm NA} \la 500$~MeV,
the effects of the nuclear medium on $\tau$ become significant
\cite{KMT,Ray92,Arellano},
and the Brueckner $g$-matrix has been used for NA and PA
scattering \cite{Satchler,M3Y,JLM,Brieva-Rook,CEG,Rikus,Amos}.

The $t$ and $g$ matrices are nonlocal operators.
In many of their applications to the analysis of experimental data, however,
they have been given as local potentials in coordinate representation.
Love and Franey \cite{Love-Franey} 
presented the $t$ matrix elements 
in the form of superposition of Yukawa potentials
that reproduce on-shell elements of the $t$ matrix.
In many of the $g$ matrix applications, it is calculated for nuclear matter,
parameterized in the form of the superposition of Yukawa or Gaussian
potentials that reproduces on-shell or half off-shell elements
of the $g$ matrix, and
translated to finite nuclei by a local density (LD) approximation.
Although the justification of the LD approximation remains as a fundamental
question, the use of $g$ as $\tau$ is successful
in reproducing the NA scattering over a wide energy range of
$50 \la E_{\rm NA} \la 800$~MeV \cite{Amos}.
For PA scattering, however, there is an ambiguity
in the definition of LD \cite{Sakuragi}. Although further progress
is required on this point, the $g$ matrix may be one of the
most plausible substitutes for $\tau$.

The local potentials thus obtained are ambiguous in their
$r_{ij}$ dependence particularly at small $r_{ij}$.
Actually, the Yukawa-type local potential is singular at $r_{ij}=0$, but
the Gaussian-type is not. Obviously, the singularity is an artifact.
Since the accuracy of the eikonal approximation is
sensitive to the dependence, the localized potentials mentioned above cannot
be used in the present theory.

In order to determine a local $\tau$ matrix from a nonlocal one
$\tau(\vrr_{ij},\vrr_{ij}')$
with no ambiguity, we introduce
the trivially equivalent local (TEL) $\tau$ matrix
$\tau^{\rm loc}(\vrr_{ij})$ as
\beq
v(\vrr_{ij})\psi
=\tau^{\rm loc}(\vrr_{ij})\phi_{0}
\label{equiv-tau}
\eeq
with $\phi_{0}=(2\pi)^{-3/2} \exp(i \vk_{ij} \cdot \vrr_{ij})$.
In the case of $\tau=t$,
$\psi$ is the solution
of the Schr\"odinger equation of the NN system with $v$,
while in the case of $\tau=g$ it is the solution of
the Bethe-Goldstone equation.
Since the discussion is parallel between the two cases,
we consider, hereafter, the case of $\tau=t$.

The TEL $\tau$ matrix, $\tau^{\rm loc}(\vrr_{ij})$,
is related to the nonlocal matrix as
\beq
\tau^{\rm loc}(\vrr_{ij})\phi_0(\vrr_{ij})
= \int d\vrr_{ij}'
\tau(\vrr_{ij},\vrr_{ij}') \phi_{0}(\vrr_{ij}') \;,
\label{equiv-tau-2}
\eeq
because the left-hand side of (\ref{equiv-tau}) is
equivalent to the right hand side of (\ref{equiv-tau-2}).
The Fourier transform of the right-hand side of (\ref{equiv-tau-2})
gives the half off-shell elements of $\tau(\vrr_{ij},\vrr_{ij}')$.
Equation (\ref{equiv-tau-2}) ensures that $\tau^{\rm loc}(\vrr_{ij})$
gives the same half off-shell elements as the nonlocal matrix
$\tau(\vrr_{ij},\vrr_{ij}')$.
Thus, $\tau^{\rm loc}(\vrr_{ij})$ is derived from the nonlocal matrix
with no ambiguity.

Even if $v(\vrr_{ij})$ is a central force,
$\tau^{\rm loc}(\vrr_{ij})$ is not a central one,
because it depends on the angle $\theta$
between $\vrr_{ij}$ and the initial momentum
$\hbar \vk_{ij}$ (the $z$ axis). This is not a problem
in the formalism of \S~\ref{Formulation}, since
the integration of $\tau^{\rm loc}(\vrr_{ij})$ over $z_{ij}$,
\beq
\tilde{\tau}^{\rm loc}(\vb_{ij})
=\int dz_{ij} \tau^{\rm loc}(\vrr_{ij}) ,
\eeq
is an input. The $\theta$ dependence of
$\tau^{\rm loc}(\vrr_{ij})$ is integrated out in
$\tilde{\tau}^{\rm loc}(\vb_{ij})$ through the integration
over $z_{ij}=r_{ij}\cos(\theta)$.

Usually, the TEL potential $\tau^{\rm loc}$ is
 a function of the relative coordinate
$\vrr_{ij}$ and $\nabla_{\vir_{ij}}$, and the spin and isospin of
the colliding pair.
When the target is a double-magic nuclus, the tensor part of
$\tau^{\rm loc}$ does not contribute to
$U^{(0)}_{\rm opt}$. For this reason,
in most analyses based on the folding model, 
only the central and spin-orbit parts
are taken into account.
Actually, the tensor force is difficult to include in the 
Glauber theory because of its noncommutative 
character \cite{Glauber}.
Thus, we assume
\bea
\tau^{\rm loc}=\tau_c+\tau_{so}\Sigma\cdot \vL  \;,
\label{tauhat}
\eea
where $\vL=-i \vrr_{ij} \times \nabla_{\vir_{ij}}$
and $\Sigma$ is the sum of the Pauli spin matrices for the two
nucleons $i$ and $j$.
The angular momentum $\hbar \vL$ can be replaced by
$\hbar \vrr_{ij} \times \vk_{ij}$ 
if $|\tau_{so}/E_{ij}| \ll 1$ and $k_{ij} a_{so} \gg 1$, 
where $a_{so}$ is the width of the region in which $\tau_{so}$
changes rapidly.
The potentials of the form (\ref{tauhat}) are 
most suited to the Glauber approximation, which we assumed 
in \S~\ref{Formulation}.

\begin{figure}[htb]
\includegraphics[width=0.5\textwidth]{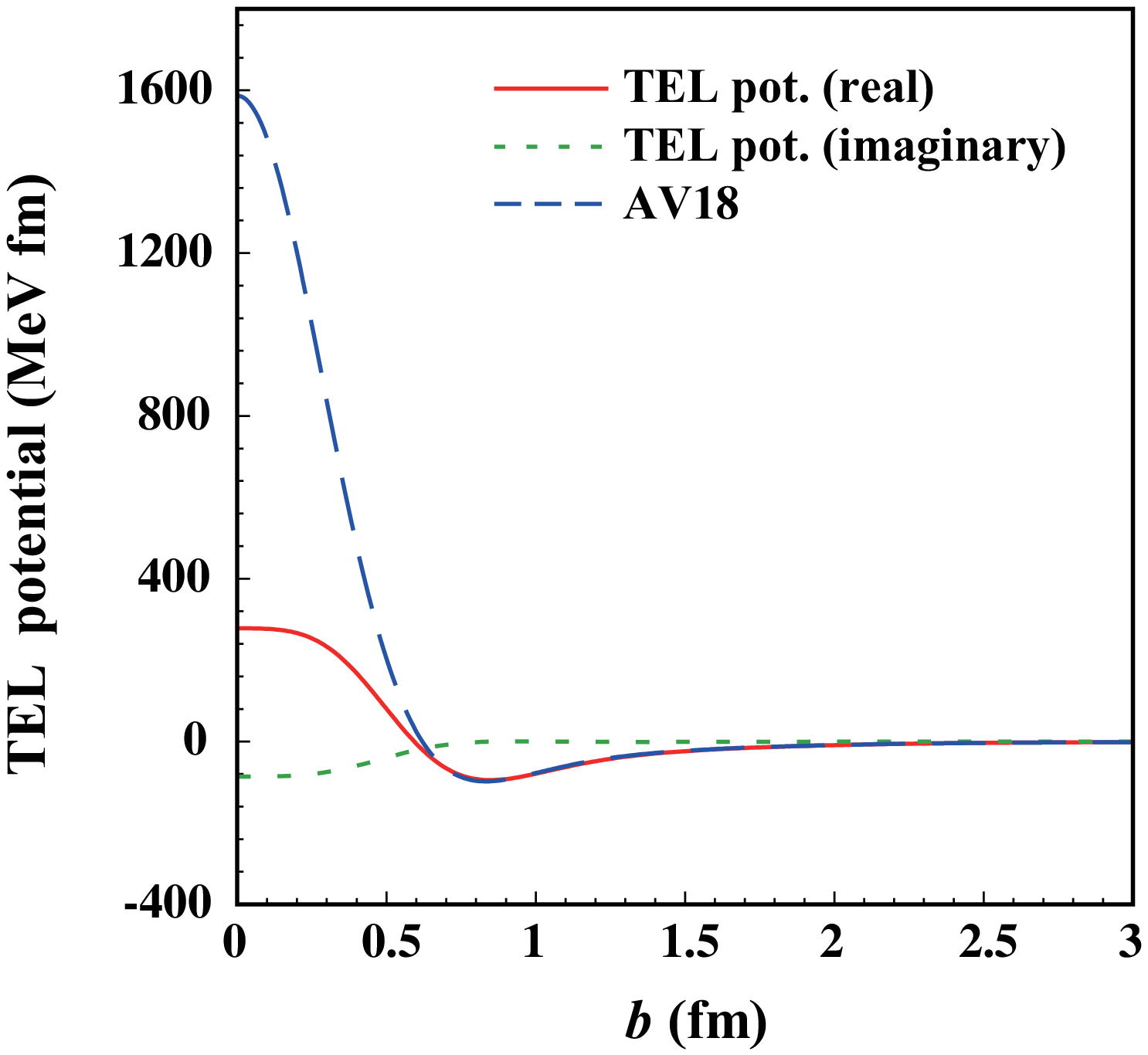}
\includegraphics[width=0.5\textwidth]{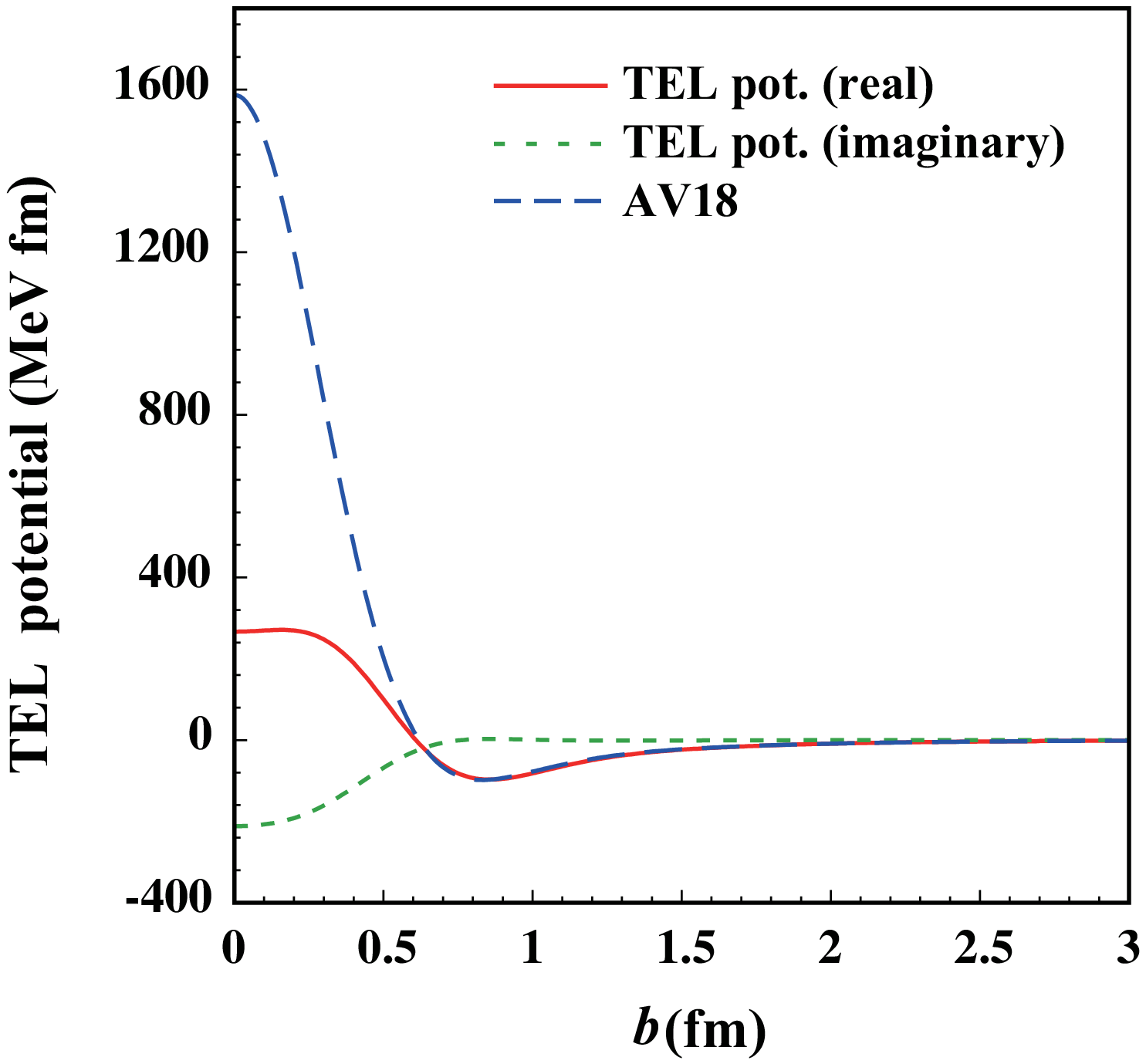}
\caption{$z_{ij}$ integrated TEL potential as a function
of $b_{ij}$ for (a) $E_{\rm NN}=300$~MeV and (b) $E_{\rm NN}=800$~MeV.
The solid (dotted) curve represents the real (imaginary) part, while
the dashed curve corresponds to the $z_{ij}$ integration of $v(\vrr_{ij})$.
}
\label{fig:tlocal}
\end{figure}

Now we calculate $\tilde{\tau}^{\rm loc}$ by solving
the Schr\"odinger equation of the NN system with $v$.
For simplicity, we neglect the spin and isospin of
the colliding pair. As $v(\vrr_{ij})$, we take
the central part of AV18 \cite{Wiringa} for the triplet-even state.
The results are shown in Fig.~\ref{fig:tlocal}, which 
represents the $z_{ij}$ integrated TEL potential
$\tilde{\tau}^{\rm loc}$ as a function
of $b_{ij}$ for the laboratory energies of 
$E_{\rm NN}=300$ and $800$~MeV.
The solid (dotted) curve is the real (imaginary) part, while
the dashed curve corresponds to $\tilde{v}(\vb_{ij})$, 
the $z_{ij}$ integration of $v(\vrr_{ij})$;
at $b_{ij}=0$, $\tilde{v}=1587~{\rm MeV} \cdot {\rm fm}$ and
$\tilde{\tau}^{\rm loc}=278-86i~(267-212i)~{\rm MeV} \cdot {\rm fm}$
for $E_{\rm NN}=300~(800)$~MeV.
As expected, $\tilde{\tau}^{\rm loc}$ has a much weaker $b_{ij}$
dependence than $\tilde{v}$. The TEL potential $\tau^{\rm loc}$
thus obtained is not singular at $b_{ij}=0$.
It is natural to think that the TEL potential of the $g$ matrix also
maintains the same property.

\subsection{Validity of eikonal and adiabatic approximations in
derivation of $\Gamma_{\rm {NN}}^{({\rm eff})}$}
\label{Validity}

The effective NN profile function
$\Gamma_{\rm {NN}}^{({\rm eff})}(\vb_{ij})$ defined by (\ref{GammaNN-eff}),
which describes a collision of the $i$-$j$ pair in the PA scattering,
is the key element of the present formalism, 
as shown in (\ref{chi-NN}) -- (\ref{GammaNN-eff}).
We test the validity of the eikonal and adiabatic approximations 
used in the derivation of $\Gamma_{\rm {NN}}^{({\rm eff})}$.

We assume $Y \gg 1$ and 
consider the accuracy of 
\bea
\Gamma_{\rm {NN}}^{({\rm eff})}(\vb_{ij}) \approx
1-{\rm exp}[-i \tilde{t}^{\rm loc}/(\hbar v_{\rm rel})] ,
\label{GammaNN-eff-2}
\eea
taking $t^{\rm loc}$ as $\tau^{\rm loc}$. 
From (\ref{chi-NN}) through (\ref{GammaNN-eff}) it is clear that 
\bea
{\cal T}_{\rm Gl} = C
\int d\vb_{ij} \; e^{i \viq \cdot \vib_{ij}} 
\Gamma_{\rm {NN}}^{({\rm eff})}(\vb_{ij}) 
\eea
is the Glauber approximation of the transition matrix of a 
fictitious PA collision in which only nucleons $i$ in P and 
$j$ in A interact. 
The exact wave function $\psi$ describing the same process satisfies 
the Schr\"odinger equation
\bea
(K+h_{\rm P}+h_{\rm A}+t^{\rm loc}(\vrr_{ij})-E)\psi=0  \;,
\label{schrodinger-3}
\eea
where $K=-(\hbar\nabla_{\viR})^2/(2\mu_{\a})$ with $\mu_{\a}$ 
the reduced mass
of PA system, $\vrr_{ij}=\vrR+\vx_{i}-\vy_{j}$ and
$E=E_{\rm PA}+e_{\rm P}+e_{\rm A}$ with
the incident energy $E_{\rm PA}$ and the intrinsic energy
$e_{\rm P}$ ($e_{\rm A}$) of P (A). 
The exact transition matrix ${\cal T}_{\rm ex}$ 
calculated using $\psi$ is 
\beq
 {\cal T}_{\rm ex}=t^{\rm loc}(\vrr_{ij})+t^{\rm loc}(\vrr_{ij})
 G_0 {\cal T}_{\rm ex} ,
\label{LS-eq-ex}
\eeq
where 
\beq
G_0=\frac{1}{E-K-h_{\rm P}-h_{\rm A}+i \epsilon} .
\eeq

First, we test the validity of the eikonal approximation. 
For this purpose, we apply the adiabatic approximation 
to Eq. (\ref{schrodinger-3}). The equation is reduced to
\bea
(K+t^{\rm loc}_{ij}(\vrr_{ij})-E_{\rm PA})\psi^{\rm AD}=0  \; ,
\label{schrodinger-4}
\eea
where the adiabatic approximation 
$h_{\rm P}+h_{\rm A}$ has been replaced by the ground state energies
$e_{\rm P}+e_{\rm A}$. We then obtain 
the transition matrix ${\cal T}_{\rm AD}$ under the adiabatic approximation as 
\beq
 {\cal T}^{\rm AD}=t^{\rm loc}(\vrr_{ij})+t^{\rm loc}(\vrr_{ij})
 G_0^{\rm AD} {\cal T}^{\rm AD} ,
\label{LS-eq-AD}
\eeq
where
\beq
G_0^{\rm AD}=\frac{1}{E_{\rm PA}-K+i \epsilon} .
\eeq
Since $\vx_{i}$ and
$\vy_{j}$ are simply parameters in (\ref{schrodinger-4}),
we can regard $K$ as $K=-(\hbar\nabla_{\vir_{ij}})^2/(2\mu_{\a})$
and then solve (\ref{schrodinger-4}) without the eikonal approximation.

Now it is possible to test the accuracy of the eikonal approximation 
by comparing ${\cal T}_{\rm Gl}$ with ${\cal T}_{\rm AD}$.
As an example, we consider $^{4}$He+$^{208}$Pb scattering.
As mentioned in \S~\ref{Effective interaction},
$t^{\rm loc}(\vrr_{ij})$ is dependent on $\theta$.
In order to avoid this complication,
we use the $\theta$-independent (spherical)
potential
\bea
t^{\rm loc}_0(r_{ij}) = - \frac{1}{\pi}\frac{1}{r_{ij}}
\frac{d}{dr_{ij}} \int_{r_{ij}}^{\infty}
\frac{\tilde{t}^{\rm loc}(b_{ij})b_{ij} db_{ij}}{(b_{ij}^2-r_{ij}^2)^{1/2}} \;.
\label{inverse-2}
\eea
instead of $t^{\rm loc}(\vrr_{ij})$.
As found from the discussion in \S~\ref{Elastic-scattering},
the two potentials $t^{\rm loc}_0(r_{ij})$ and $t^{\rm loc}(\vrr_{ij})$ 
give the same transition matrix under the eikonal approximation.

\begin{figure}[htb]
\centerline{\includegraphics[width=0.5\textwidth]{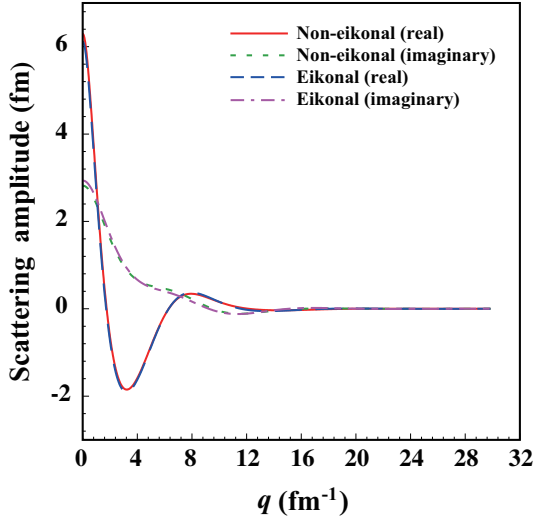}}
\caption{
On-shell scattering amplitudes, 
$f_{\rm Gl}(\vq)$ and $f_{\rm AD}(\vq)$, 
of a single NN collision in $^{4}$He+$^{208}$Pb scattering
$t^{\rm loc}_0(r_{ij})$ 
at the laboratory energy of $E_{\rm PA}=300 N_{\rm P}$ MeV. 
}
\label{fig:fcomp2}
\end{figure}

Figure \ref{fig:fcomp2} shows the scattering amplitudes, 
$f_{\rm Gl}(\vq)=-(2\pi)^2\mu /\hbar^2 \times {\cal T}_{\rm Gl}$
and 
$f_{\rm AD}(\vq)=-(2\pi)^2\mu /\hbar^2 \times {\cal T}_{\rm AD}$,
of a single NN collision in $^4$He+$^{208}$Pb scattering
at the laboratory energy of $E_{\rm PA}=300 N_{\rm P}$~MeV.
The solid (dashed) and dotted (dash-dotted) curves show, respectively,
the real and imaginary parts of $f_{\rm AD}(\vq)$ ($f_{\rm Gl}(\vq)$). 
The agreement between calculations with and without the eikonal approximation
is excellent.

The denominator of $G_0^{\rm AD}$ in the momentum representation is
\beq
\frac{\hbar^2(\vk^2-\vk'^2)}{2\mu_{\a}}
=\hbar v_{\rm rel}\Big(q_z-\frac{\vq^2\hbar}{2\mu_{\a} v_{\rm rel}}\Big),
\label{denominator1}
\eeq
where $v_{\rm rel}=\hbar k/\mu_{\a}$, $E_{\rm PA}=\hbar^2 \vk^2/2\mu_{\a}$,
$\vq=\vk-\vk'$ and $q_z$ is the $z$ component of $\vq$.
In the eikonal approximation, the $\vq^2$ term in the denominator 
is neglected.
This is realized in the large-$\mu_{\a}$ limit with $v_{\rm rel}$ fixed.
In general, PA scattering has a larger $\mu_{\a}$
than NA scattering,
and $\mu_{\a}$ becomes minimum for NN scattering.
Thus, the eikonal approximation is generally better for PA scattering
than for NA scattering but becomes worse for NN scattering,
when the scatterings of different systems with common $v_{\rm rel}$
are compared with each other.

In order to confirm the validity of the adiabatic approximation, we compare
the solution ${\cal T}_{\rm ex}$ of Eq.~(\ref{LS-eq-ex}) with 
${\cal T}^{\rm AD}$
of Eq.~(\ref{LS-eq-AD}) for nucleus-proton scattering in which
the nucleus is described by a core+nucleon ($c+n$) two-body system. 
We consider the two reactions $^{11}$Be$+p$ and $^{40}$Ca$+p$
at an incident energy of 300 MeV per nucleon.
We use a Woods-Saxon interaction between $c$ ($^{10}$Be or $^{39}$Ca)
and $n$
\beq
V_{cn}(r)=V_0 \left(1+\exp\frac{r-R_0}{a_0}\right)^{-1},
\label{WSpot}
\eeq
where $r$ is the displacement of $n$ from $c$. The potential
parameters and the resulting neutron separation energy $S_n$
are shown in Table \ref{tab-pot}.
\begin{table}[hptb]
\caption{\label{tab-pot}
Potential parameters in Eq.~(\ref{WSpot}) and neutron separation
energy $S_n$ for $^{10}$Be-$n$ and $^{39}$Ca-$n$.
}
\begin{center}
\begin{tabular}{ccccc} \hline \hline
              & $V_0$ (MeV) & $R_0$ (fm) & $a_0$ (fm) & $S_n$ (MeV) \\ \cline{2-5}
$^{10}$Be-$n$ & 51.60 & 2.996 & 0.52 & 0.503 \\ \hline
$^{39}$Ca-$n$ & 51.68 & 4.343 & 0.67 & 15.64 \\ \hline
\end{tabular}
\end{center}
\end{table}
We solve the three-body scattering problem by 
the method of continuum-discretized coupled
channels (CDCC) \cite{CDCC,CDCC-foundation} 
with and without 
adiabatic approximation to the $c$-$n$ internal Hamiltonian, i.e.,
$h_{\rm P}$ in Eq.~(\ref{schrodinger-3}). For simplicity, we neglect 
the internal degrees of freedom of the target proton; we then have
$h_{\rm A}=0$ in the present case. The maximum $r$ is 60 fm, and
$s$-, $p$-, $d$- and $f$-waves for the $c$-$n$ relative wave function
are included. The momentum bin is truncated at $1.5$ fm$^{-1}$
for each partial wave and divided into 30 (15) discretized
states for the $s$-wave ($p$-, $d$- and $f$-waves). The coupled-channel
potentials are calculated using $t^{\rm loc}_0(r_{ij})$ of
Eq.~(\ref{inverse-2}). Note that the interaction between $c$ and the
target proton is not included since we solve the Schr\"{o}dinger
equations (\ref{schrodinger-3}) and (\ref{schrodinger-4}).
The relative wave functions $\psi$ and $\psi^{\rm AD}$ are
integrated up to 20 fm and connected to the standard asymptotic form.

\begin{figure}[htb]
\includegraphics[width=0.5\textwidth]{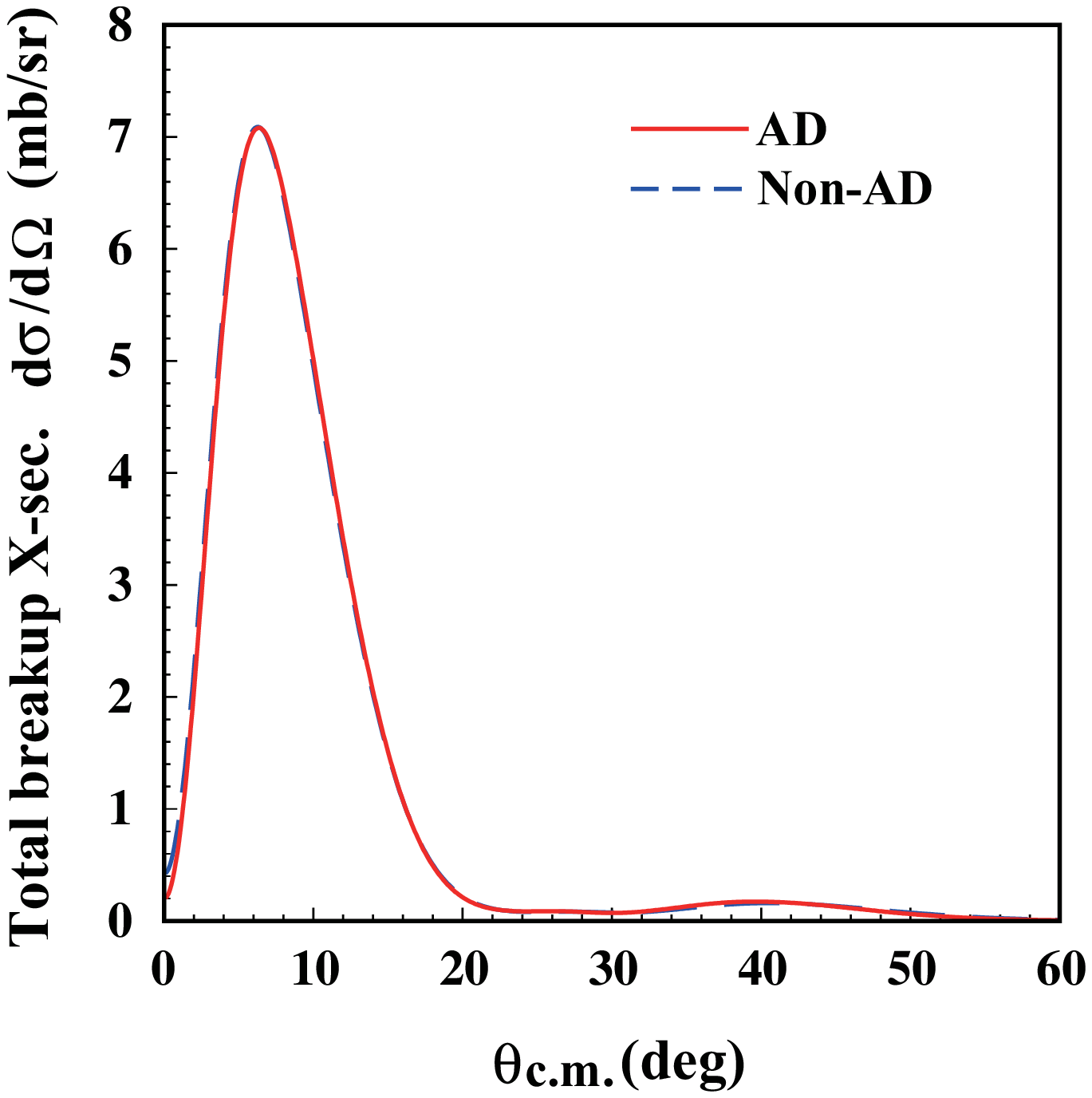}
\includegraphics[width=0.5\textwidth]{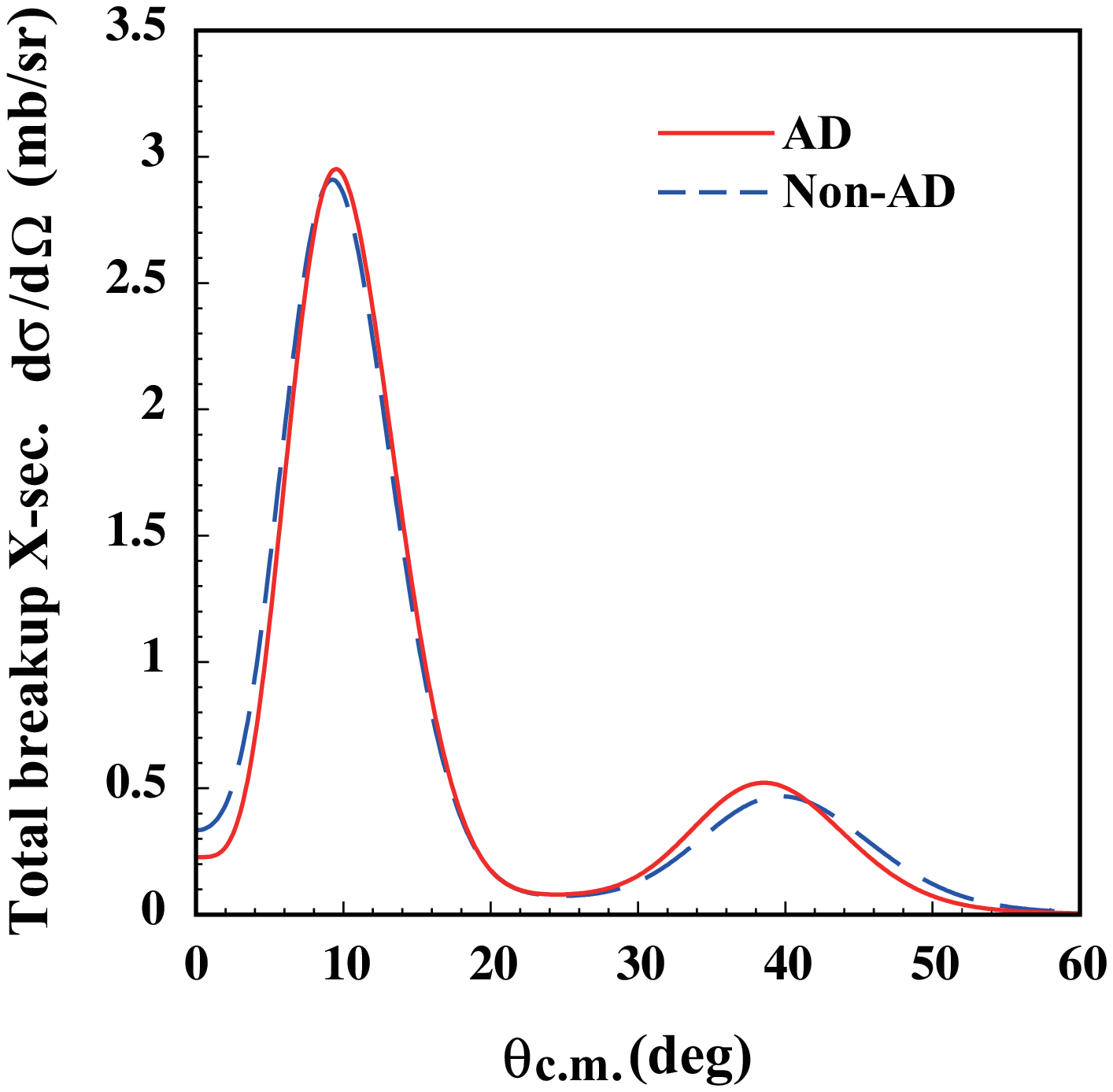}
\caption{
Total breakup cross sections for
$^{11}$Be$+p$ (left panel) and $^{40}$Ca$+p$ (right panel)
at 300 MeV per nucleon. The horizontal axis is the
scattering angle of the center of mass of the fragments
$c$ and $n$.
The solid and dashed lines show the results of CDCC
calculation with and without adiabatic approximation, 
respectively.
}
\label{fig:CDCC}
\end{figure}
In Fig.~\ref{fig:CDCC} we show the total breakup cross sections
for $^{11}$Be$+p$ (left panel) and $^{40}$Ca$+p$ (right panel)
at 300 MeV per nucleon as a function of the center-of-mass
scattering angle of the $c$-$n$ two-body system. In each panel
the solid and dashed lines indicate the results with and without
adiabatic approximation, respectively. One sees clearly that the
adiabatic approximation works very well for not only $^{11}$Be
with small $S_n$ but also $^{40}$Ca with quite large $S_n$.
Thus, in the energy region considered here, namely, a few 
hundred MeV per nucleon, the adiabatic approximation and 
hence Eq.~(\ref{schrodinger-4}) is shown to be valid.

\subsection{Relationship between present theory and 
conventional Glauber calculations}
\label{Relation}

In this subsection, we clarify the relationship
between the present theory and the conventional Glauber calculations using 
the empirical NN profile function  (\ref{EqMK1-5}) \cite{GM}
or using the modified profile functions including 
the in-medium effects \cite{Xiang}.
For simplicity,  we consider the case of $Y \gg 1$, and
the superscript loc of $\tau^{\rm loc}$ is suppressed.

In the present theory, the transition
matrix of the PA scattering in coordinate
representation is
\bea
T=C
\Big[
1-\prod_{i,j}
\exp(-i {\tilde \tau}_{ij}/\hbar v_{\rm rel} )
\Big]
\;.
 \label{T-present}
\eea
The corresponding matrix in the Glauber theory is
\bea
 T_{\rm Gl}
 =C
\Big[
 1-\prod_{i,j}(1-i {\tilde t^{\rm em}_{ij}}/\hbar v_{\rm rel})
\Big]
 \;,
 \label{T-Gl}
\eea
where ${\tilde t^{\rm em}_{ij}}=-i\hbar v_{\rm rel}
\Gamma^{\rm em}_{\rm NN}$ using $\Gamma^{\rm em}_{\rm NN}$ of (\ref{EqMK1-5}).
Since ${\tilde t^{\rm em}_{ij}}$ is adjusted to fit
the data on NN scattering, it is
essentially identical to ${\tilde t_{ij}}$.
One sees, therefore, that the conventional Glauber calculation with
(\ref{EqMK1-5}) is justified provided that (a) $\tau \approx t$
is a good approximation and (b) ${\tilde \tau}_{ij}/\hbar v_{\rm rel}$
is small enough to warrant
$\exp(-i {\tilde \tau}_{ij}/\hbar v_{\rm rel} ) \approx
1 - i {\tilde \tau}_{ij}/\hbar v_{\rm rel}$.
At very high energies, where the replacement of $\tau$ by $t$ is valid and
the amount of multiple scatterings by $t$ is negligible,
that is, in the limit that the impulse approximation is good,
the present theory agrees using the conventional Glauber theory
with the empirical profile function $\Gamma^{\rm em}_{\rm {NN}}$
of (\ref{EqMK1-5}). Conversely,
if either of conditions (a) and (b)
is not satisfied, the conventional Glauber procedure is not justified
by the present approach.

In the heuristic approach in Ref. \citen{Xiang},
${\tilde t^{\rm em}_{ij}}$ is replaced by 
the $z$ integration of $g_{ij}$, ${\tilde g_{ij}}$, 
in the conventional Glauber theory.
The corresponding transition matrix of the PA scattering is
\bea
 T^{\rm (g)}
 =C
\Big[
 1-\prod_{i,j}(1-i {\tilde g_{ij}}/\hbar v_{\rm rel})
\Big]
 \;.
\label{T-g}
\eea
One can find from (\ref{T-present}) and (\ref{T-g})
that even if ${\tilde \tau_{ij}}\approx{\tilde g_{ij}}$,
$T^{\rm (g)}$ agrees with $T$ in the lowest (first) order
of ${\tilde g}/\hbar v_{\rm rel}$, but not in higher orders.
Thus, the heuristic approach in Ref. \citen{Xiang}
gives a better description of PA scattering than
the Glauber theory, but it is still not perfect,
since it contains no higher order terms of
${\tilde g}/\hbar v_{\rm rel}$.

For the calculation of optical potential, it is common to take
the optical limit 
in which only the first order of
${\tilde g}/\hbar v_{\rm rel}$ is taken into account. In this limit,
the present theory with $g$ as $\tau$ reduces to
the folding model with $g$, which has been successfully used in reproducing
NA  scattering data \cite{Rikus,Amos}.

\section{Summary}
\label{Summary}

We present in this paper a new version of the Glauber theory
for nucleus-nucleus collisions based on the multiple scattering theory
of Watson (MST) \cite{Watson},
making use of the KMT formalism~\cite{KMT}.
The input of the theory is the effective nucleon-nucleon (NN) interaction
$\tau$ of MST, which has a weaker short-range repulsion
than the bare NN nuclear force potential,
which makes the eikonal approximation much more reliable.
We tested the validity of the eikonal and adiabatic approximations
to NN collision using $\tau$ in nucleus-nucleus
scattering at the laboratory energy of 
$E_{\rm PA}=300 N_{\rm P}$~MeV and showed that these approximations are
good at the intermediate energies.

At very high energies where the replacement of $\tau$ by $t$ is valid and
multiple scattering by $t$ is negligible,
that is, in the limit that the impulse approximation is good,
the present theory gives the same $T$ matrix of PA scattering as
the conventional Glauber theory using the empirical profile function
$\Gamma^{\rm em}_{\rm {NN}}$ of (\ref{EqMK1-5}).

When the $g$ matrix is used as $\tau$, the present theory
can take into account the nuclear-medium effects in the analyses of
NA and PA scattering. Therefore, the present theory is also 
applicable in the intermediate-energy region.
The present theory also
provides a theoretical foundation to
heuristic Glauber calculations \cite{Xiang}
in which the profile function is modified to reproduce the
in-medium NN cross section calculated from the $g$ matrix,
although the heuristic calculations include no multiple scattering by $g$.
For the calculation of optical potential,
the present formalism reduces in the optical limit to
the folding model using $g$ matrix, which has been successfully used 
in reproducing nucleon-nucleus scattering data \cite{Rikus,Amos}.

Thus, the present theory unifies standard methods such as
the conventional Glauber method \cite{Glauber} using the empirical NN
profile function (\ref{EqMK1-5}) \cite{GM},
heuristic Glauber calculations with the modified profile functions
including the in-medium NN effects \cite{Xiang},
and the folding potential method using the $g$ matrix
\cite{Satchler,M3Y,JLM,Brieva-Rook,CEG,Rikus,Amos}.
In our forthcoming paper, we propose a
way of deriving $\tau$ with no local density approximation,
and compare the result with those obtained using the $t$ and $g$ matrices.

\section*{Acknowledgements}
This work has been supported in part by a Grant-in-Aid
for Scientific Research (18540280)
from the Ministry of Education, Culture, Sports, Science and Technology 
of Japan.




\begin{thebibliography}{99}

\bibitem{Tanihata}
        I. Tanihata et al., Phys. Rev. Lett. {\bf 55} (1985), 2676;
        Phys. Lett. B {\bf 287} (1992), 307.

\bibitem{Glauber}
        R. J. Glauber, {\it Lectures in Theoretical
        Physics} (Interscience, New York, 1959), Vol. 1, p. 315.


\bibitem{Tostevin}
J. S. Al-Khalili and J. A. Tostevin, Phys. Rev. Lett. {\bf 76} (1996),
3903.

J. A. Tostevin and B. A. Brown, Phys. Rev. C {\bf 74} (2006), 064604,
and references cited therein.

\bibitem{Ogawa01}
Y. Ogawa, T. Kido, K. Yabana and Y. Suzuki,
Prog. Theor. Phys. Suppl. No. 142 (2001), 157,
and references cited therein.




\bibitem{Xiang}
C. Xiangzhou, F. Jun, S. Wenqing, M. Yugang, W. Jiansong and
Y. Wei, Phys. Rev. C {\bf 58} (1998), 572.

A. de Vismes, P. Roussel-Chomaz and F. Carstoiu,
Phys. Rev. C {\bf 62} (2000), 064612.

R. E. Warner, I. J. Thompson and J. A. Tostevin,
Phys. Rev. C {\bf 65} (2002), 044617.


\bibitem{Horiuchi:2006ga}
  W. Horiuchi, Y. Suzuki, B. Abu-Ibrahim and A. Kohama,
  Phys. Rev. C {\bf 75} (2007), 044607.



\bibitem{GM}
R. J. Glauber and G. Matthiae, Nucl. Phys. B {\bf 21} (1970), 135.



\bibitem{Wiringa}
R. B. Wiringa, V. G. J. Stoks and R. Schiavilla,
Phys. Rev. C {\bf 51} (1995), 38.



\bibitem{Wallace}
S. J. Wallace, Ann. of Phys. {\bf 78} (1973), 190;
Phys. Rev. D {\bf 8} (1973), 1934, and references therein.



\bibitem{Wong}
C. W. Wong and S. K. Young, Phys. Rev. C {\bf 12} (1975), 1301.

S. K. Young and C. W. Wong, Phys. Rev. C {\bf 15} (1977), 2146;
Phys. Lett. B {\bf 77} (1978), 41.



\bibitem{Wallace-review}
S. J. Wallace, Adv. Nucl. Phys. {\bf 12} (1981), 135, 
and references cited therein.


\bibitem{Watson} K. M. Watson, Phys. Rev. {\bf 89} (1953), 575.


\bibitem{KMT} A. K. Kerman, H. McManus and R. M. Thaler, Ann. of 
Phys. {\bf 8} (1959), 551.


\bibitem{Glauber-foundation}
N. M. Queen, Nucl. Phys. {\bf 55} (1964), 177.

D. R. Harrington, Phys. Rev. {\bf 184} (1969), 1745.

J. M. Eisenberg, Ann. of Phys. {\bf 71} (1972), 542.

V. B. Mandelzweig and S. J. Wallace, Phys. Rev. C {\bf 25} (1982), 61.




\bibitem{Takeda}
G. Takeda and K. M. Watson, Phys. Rev. {\bf 97} (1955), 1336.

\bibitem{Picklesimer}
A. Picklesimer and R. M. Thaler, Phys. Rev. C {\bf 23} (1981), 42.

\bibitem{Love-Franey}
W. G. Love and M. A. Franey,
Phys. Rev. C {\bf 24} (1981), 1073

M. A. Franey and W. G. Love,
Phys. Rev. C {\bf 31} (1985), 488.



\bibitem{Ray92}
L. Ray, Phys. Rev. C {\bf 41} (1990), 2816.

L. Ray, G. W. Hoffmann and W. R. Coker, Phys. Rep. {\bf 212} (1992), 223,
and references cited therein.


\bibitem{Arellano}
H. F. Arellano, F. A. Brieva and W. G. Love, Phys. Rev. C
{\bf 52} (1995), 301.



\bibitem{Satchler}
G. R. Satchler, {\it Direct Nuclear Reactions}
(Oxford University Press, New York, 1983),
and references cited therein.

\bibitem{M3Y}
G. Bertsch, J. Borysowicz, H. McManus and W. G. Love,
Nucl. Phys. A {\bf 284} (1977), 399.

\bibitem{JLM}
J.-P. Jeukenne, A. Lejeune and C. Mahaux, Phys. Rev. C {\bf 16} (1977),
80; Phys. Rep. {\bf 25} (1976), 83.

\bibitem{Brieva-Rook}
F. A. Brieva and J. R. Rook, Nucl. Phys. A {\bf 291} (1977), 299;
ibid. {\bf 291} (1977), 317; ibid. {\bf 297} (1978), 206.

\bibitem{CEG}
N. Yamaguchi, S. Nagata and T. Matsuda, Prog. Theor.
Phys. {\bf 70} (1983), 459.

N. Yamaguchi, S. Nagata and J. Michiyama,
Prog. Theor. Phys. {\bf 76} (1986), 1289.

\bibitem{Rikus}
L. Rikus, K. Nakano and H. V. von Geramb, Nucl. Phys. A {\bf 414} 
(1984), 413.

L. Rikus and H. V. von Geramb, Nucl. Phys. A {\bf 426} (1984), 496.

\bibitem{Amos}
K. Amos, P. J. Dortmans, H. V. von Geramb, S. Karataglidis
and J. Raynal, Adv. Nucl. Phys. {\bf 25} (2000), 275.


\bibitem{Sakuragi}
T. Furumoto and Y. Sakuragi,
Phys. Rev. C {\bf 74} (2006), 034606.


\bibitem{CDCC}
M. Kamimura, M. Yahiro, Y. Iseri, Y. Sakuragi, H. Kameyama and M. Kawai,
Prog. Theor. Phys. Suppl. No. 89 (1986).

N. Austern, Y. Iseri, M. Kamimura, M. Kawai, G. H. Rawitscher 
and M. Yahiro, Phys. Rep. {\bf 154} (1987), 125.


\bibitem{CDCC-foundation}
N. Austern et al.,  Phys. Rev. Lett. {\bf 63} (1989),
2649; Phys. Rev. C {\bf 53} (1996), 314.






\end{thebibliography}
\end{document}